\documentclass[
    ,final            
  ]
  {aipproc}

\layoutstyle{6x9}

\begin{document}
\vspace*{-1cm}\hspace*{11cm}{\small MTA-PHYS-0606}\vspace*{0.2cm}
\title{Constraints on mSUGRA from isospin asymmetry in $B \to K^* \gamma$}

\classification{11.30.Pb, 12.15.Mm, 12.60.Jv, 13.20.He}
\keywords      {isospin asymmetry; supersymmetry; Wilson coefficients; minimal flavor violation}

\author{F. Mahmoudi}{}

\author{M. R. Ahmady}{
  address={Department of Physics, Mount Allison University, 67 York Street, Sackville,\\ New Brunswick,
 Canada E4L 1E6}
}

\begin{abstract}
We study the isospin asymmetry as an observable in exclusive $B \to K^* \gamma$ decay that can be used to study the effects of physics beyond the Standard Model. We present predictions for isospin asymmetry in Supersymmetric extension of the Standard Model with minimal flavor violation, using parameter values allowed by current experimental constraints. This observable can provide very tight constraints on the mSUGRA parameter space.
\end{abstract}

\maketitle

%
\section{Introduction}
The next generation of colliders, and especially the LHC, will boost the search for new physics, and in particular, the Supersymmetry (SUSY). Therefore, it seems important to derive new constraints on the properties of supersymmetric particles. For this purpose, a good strategy would be to look at where the Standard Model (SM) can be subdominant in comparison to the supersymmetry, and where QCD corrections are known with high accuracy and branching fractions can be measured at LHC even at initial low luminosities. Rare B decays appear to be {\it ideal choices} for the above goal. For example, $b \to s \, \gamma$ transitions are forbidden at the tree level in the SM. In this work, we will concentrate on the related exclusive $B \to K^* \gamma$ decay, within supersymmetry, considering mSUGRA parameter space and the minimal flavor violation. The main focus is on isospin asymmetry which, as we will see later, is a valuable observable to get constraints on supersymmetric parameters. A more detailed analysis is presented in Ref. \cite{mahmoudi}.

\section{Framework}

The isospin asymmetry for the exclusive process $B\to K^*\gamma$ can be defined as:
\begin{equation}
\Delta_{0-}=\frac{\Gamma (\bar B^0\to\bar K^{*0}\gamma ) -\Gamma (B^-\to K^{*-}\gamma )}{\Gamma (\bar B^0\to\bar K^{*0}\gamma )+\Gamma (B^-\to K^{*-}\gamma
)}\;\; , \label{isospinasym}
\end{equation}
and $\Delta_{0+}$ can be obtained by using the charge conjugate modes. 
We consider here recent data for exclusive decays from Belle \cite{belle} and Babar \cite{babar}:
\begin{eqnarray}
\Delta_{0-}&=& +0.050 \pm 0.045({\rm stat.})\pm 0.028({\rm syst.})\pm 0.024(R^{+/0})\;\;\; (\mbox{Babar})\; , \label{babar}\\
\Delta_{0+}&=& +0.012 \pm 0.044({\rm stat.})\pm 0.026({\rm syst.})\;\;\; (\mbox{Belle})\;. \label{belle}
\end{eqnarray}
The effective Hamiltonian for $b\to s\gamma$ transitions can be written:
\begin{equation}
{\cal H}_{eff}=\frac{G_F}{\sqrt{2}}\sum_{p=u,c} V^*_{ps}V_{pb} \left[ C_1(\mu) \, O^p_1(\mu) + C_2(\mu) \, O^p_2(\mu) +\sum_{i=3}^8 C_i(\mu) \, O_i(\mu)\right]\;\;,
\end{equation}
where $G_F$ is the Fermi coupling constant, $V_{ij}$ are elements of the CKM matrix, $O_i(\mu)$ are the operators relevant to $B\to K^*\gamma$ and $C_i(\mu)$ are the corresponding Wilson coefficients evaluated at the scale $\mu$. The most relevant operators $O_i$ in supersymmetry are:
\begin{equation}
O_7 =\frac{e}{4\pi^2} m_b \bar s_\alpha\sigma^{\mu\nu} P_R b_\alpha F_{\mu\nu} \;\; , \;\;O_8 =\frac{g_s}{4\pi^2} m_b \bar s_\alpha\sigma^{\mu\nu} P_R T^a_{\alpha\beta}b_\beta G_{\mu\nu}^a\;\;, \label{operators}
\end{equation}
where $P_L(P_R)=(1-(+)\gamma_5)/2$ are the projection operators.\\
We consider the Wilson coefficients at the next to leading order (NLO) in the strong coupling constant $\alpha_s$. The contributions from the W boson, the charged Higgs and the charginos, as well as the leading $\tan\beta$ corrections to the W boson and the charged Higgs are considered. The contributions from the gluino and neutralino are neglected in our work, as they are known to be negligible in the minimal flavor violating scenario~\cite{bertolini}. For further details of the calculation of the Wilson coefficients, please refer to \cite{mahmoudi}. Following the method of Ref. \cite{kagan}, one can write the isospin asymmetry as:
\begin{equation}
\Delta_{0-}=\mbox{Re} (b_d-b_u)\;\; .
\label{asymb}
\end{equation}
The expression for $b_q$, which is related to the Wilson coefficients, can be found in \cite{kagan}.\\
We generate the SUSY mass spectrum using ISAJET-7.74 \cite{baer} and we perform scans in the mSUGRA parameter space ($m_0$, $m_{1/2}$, $A_0$, $\mbox{sign}(\mu)$, $\tan\beta$). For any mSUGRA parameter space point, we then calculate the isospin asymmetry, and compare it to the combined experimental limits of eq.(\ref{babar}) and eq.(\ref{belle}). For comparison, we also perform the calculation of the inclusive branching ratio of $B\to X_s\gamma$ following Ref. \cite{kagan99} and compare to the experimental limits of \cite{battaglia}. The points which result in too small light Higgs masses ({\it i.e.} such as $m_{h^0} < 111$ GeV) or which do not satisfy the constraints of Ref. \cite{PDG2004} are also excluded. Finally, we also test whether the lightest supersymmetric particle (LSP) is charged. Indeed, a charged LSP is cosmologically disfavored if the LSP is stable, but might be viable if R-parity is violated.

\section{Results}
\begin{figure}[!t]
\includegraphics[width=8cm,height=6.15cm]{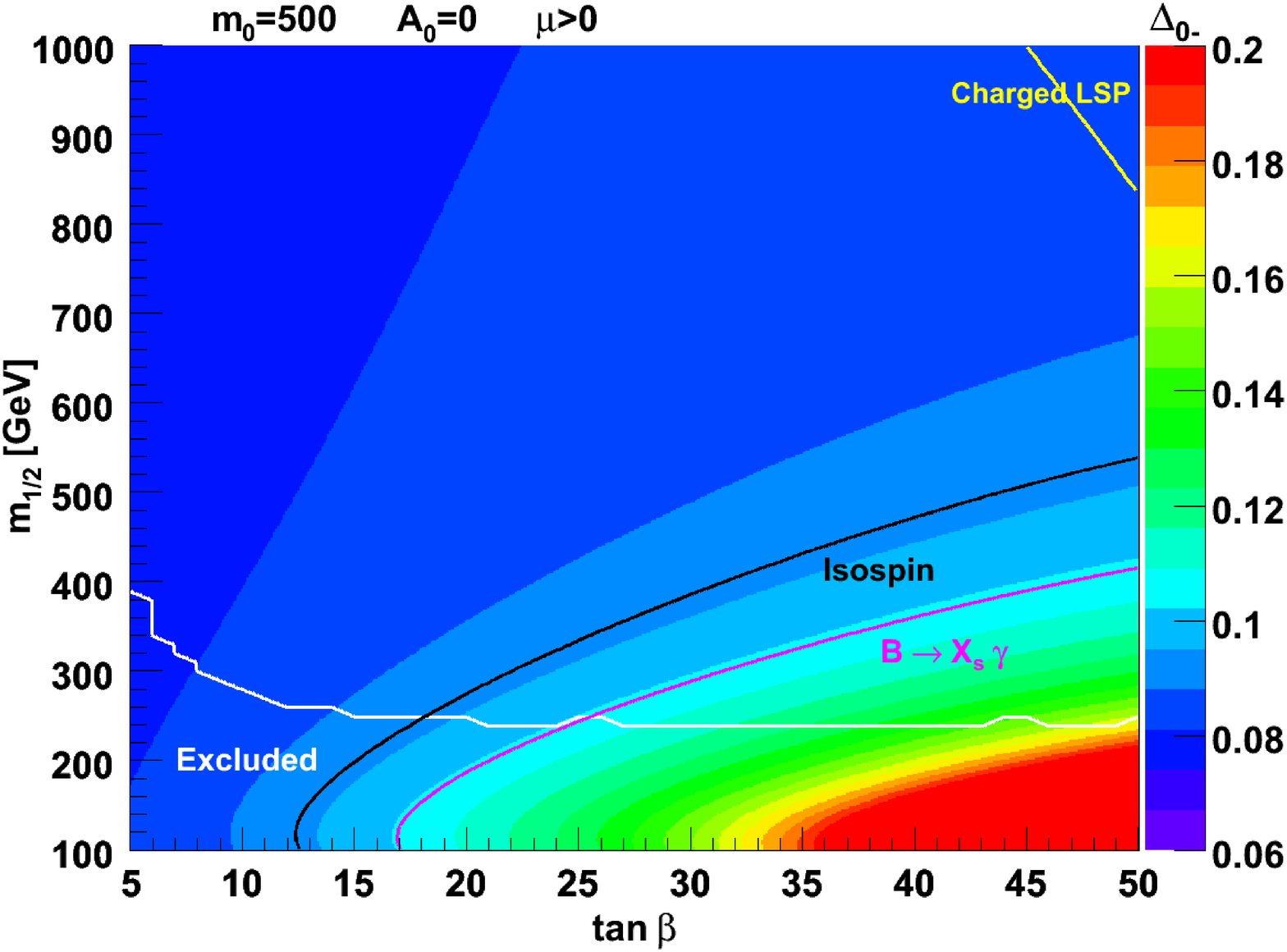}~\includegraphics[width=8cm,height=6.15cm]{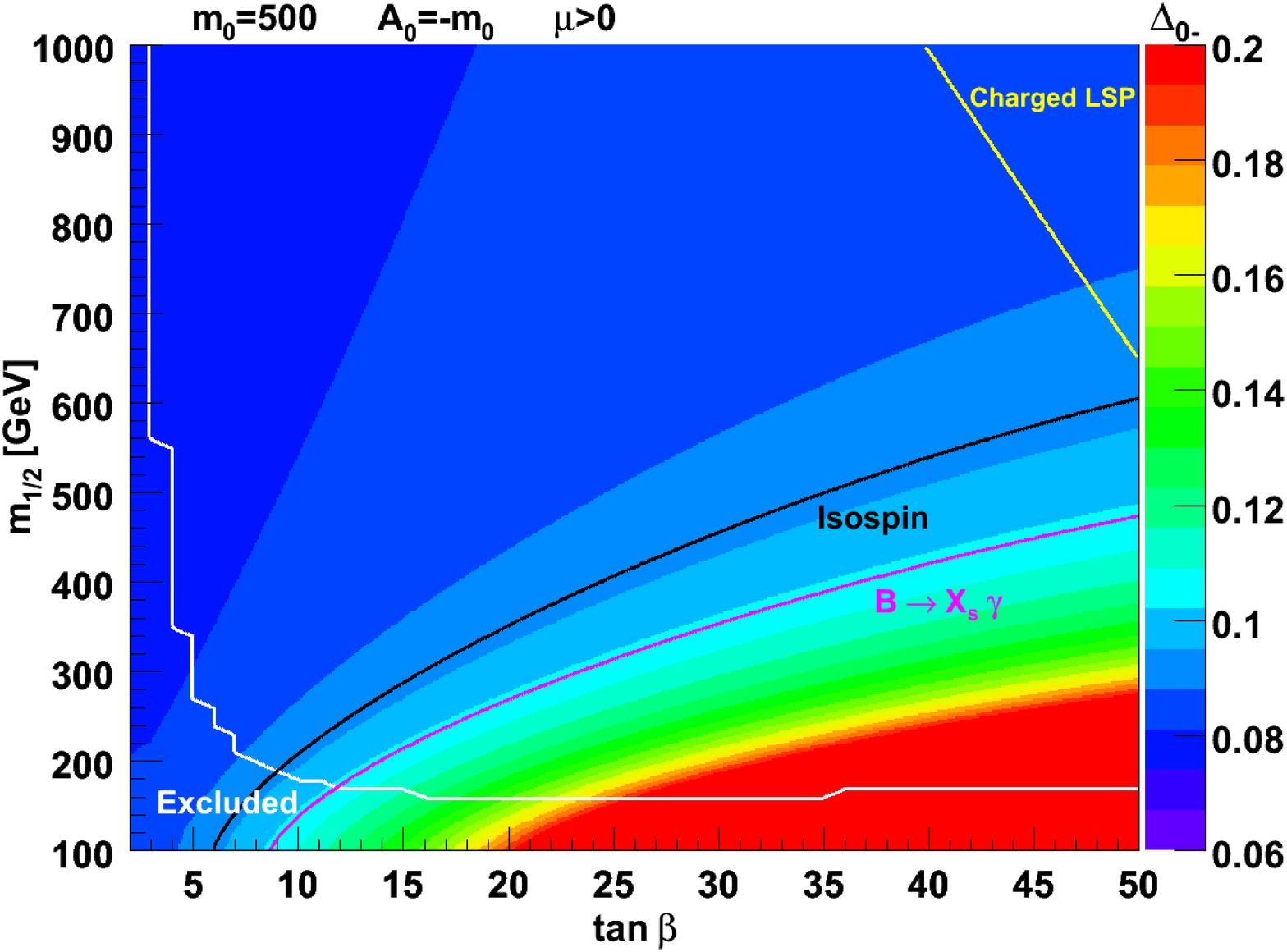}
\caption{Constraints on the mSUGRA parameter plane $(\tan\beta,m_{1/2})$ for $m_0=500$, with $A_0=0$ and $A_0=-m_0$. The definitions of the different regions are in the text.}\label{tanb}
\end{figure}

We perform scans of the mSUGRA parameter space such that $m_0 \in [0,1000]$, $m_{1/2} \in [0,1000]$, $\tan\beta \in [0,50]$, $A_0 \in [-1000,1000]$ and $\mu>0$.\\
\begin{figure}[!ht]
\begin{tabular}{cc}
\includegraphics[width=8cm,height=6.2cm]{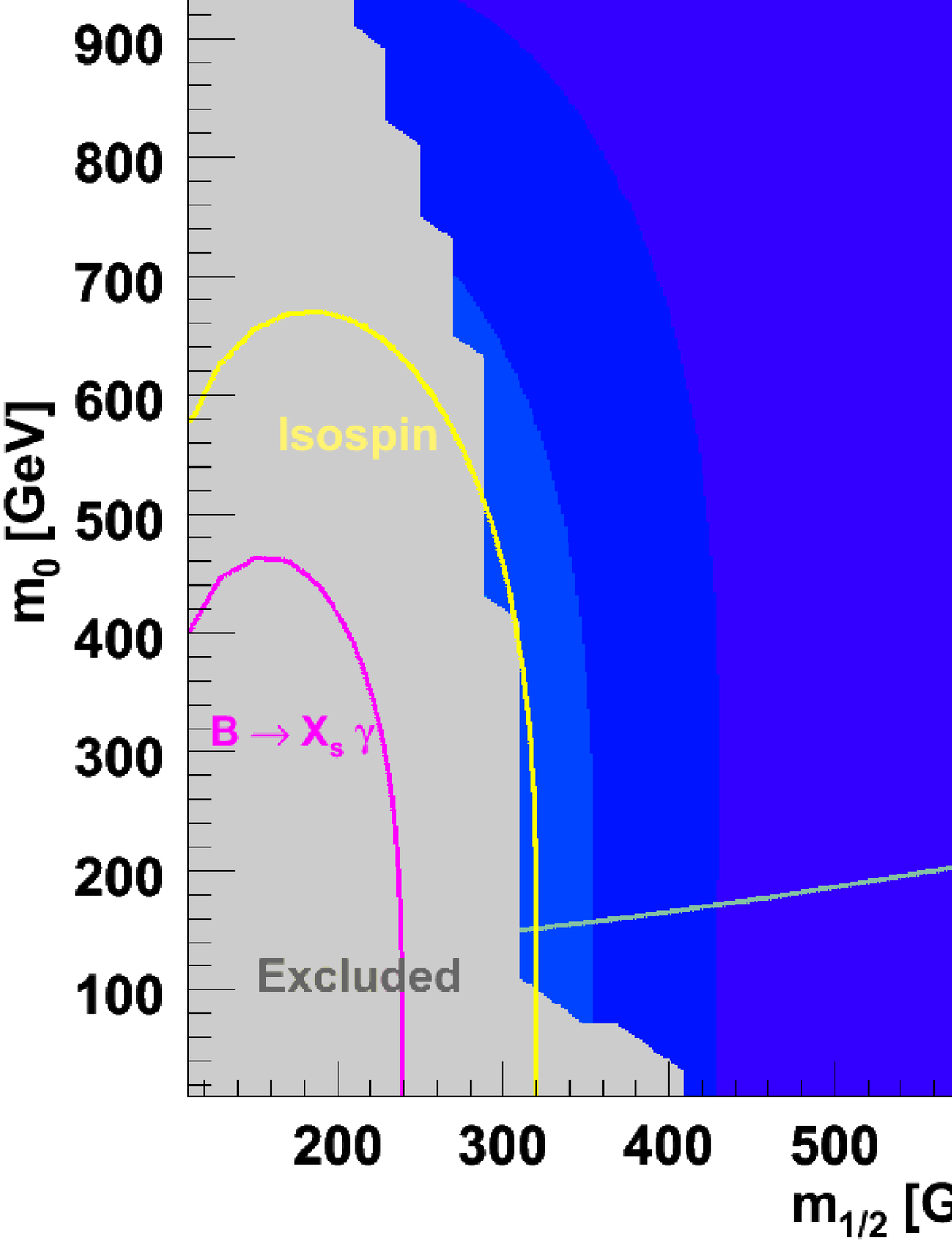}&\includegraphics[width=8cm,height=6.2cm]{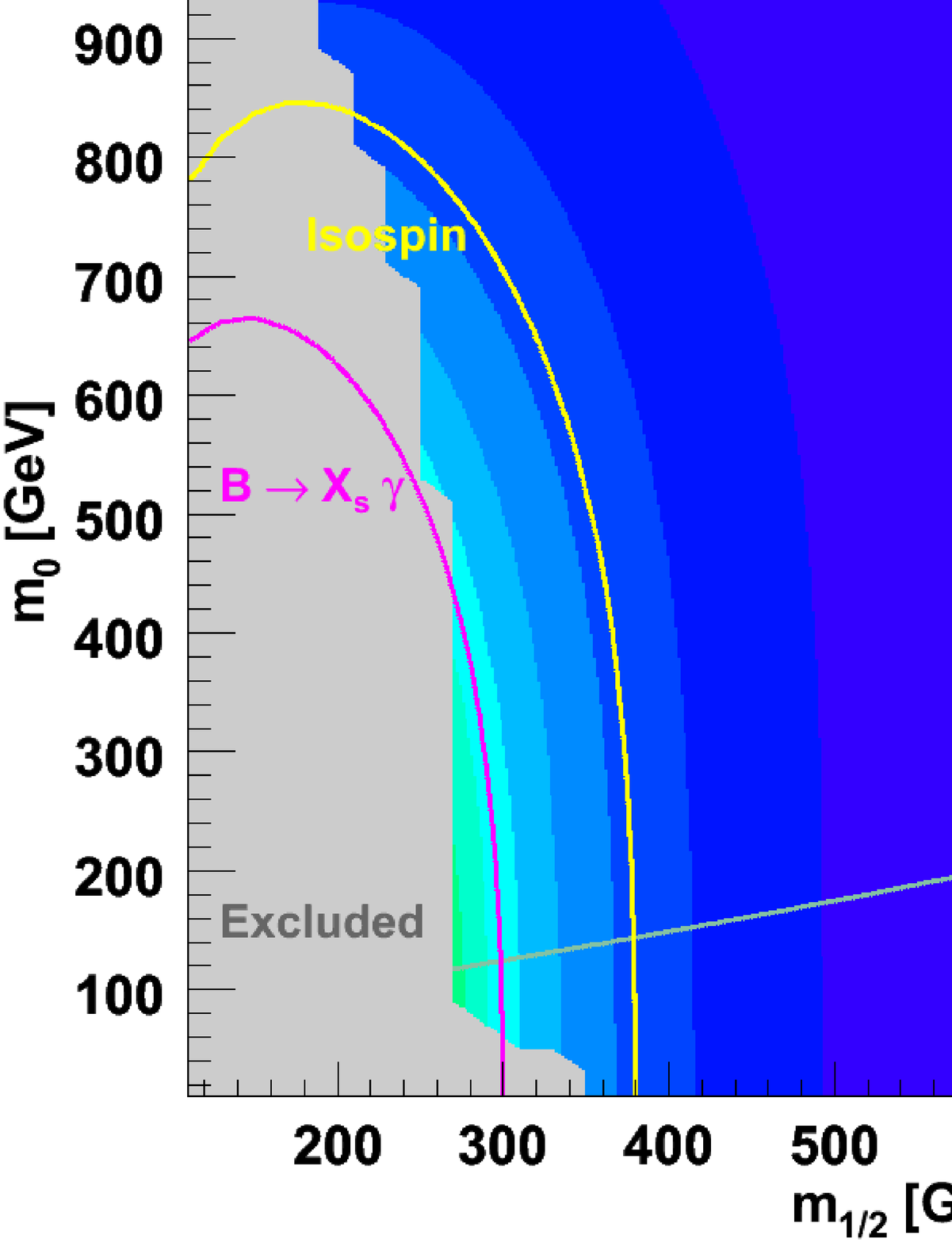}\\
\includegraphics[width=8cm,height=6.2cm]{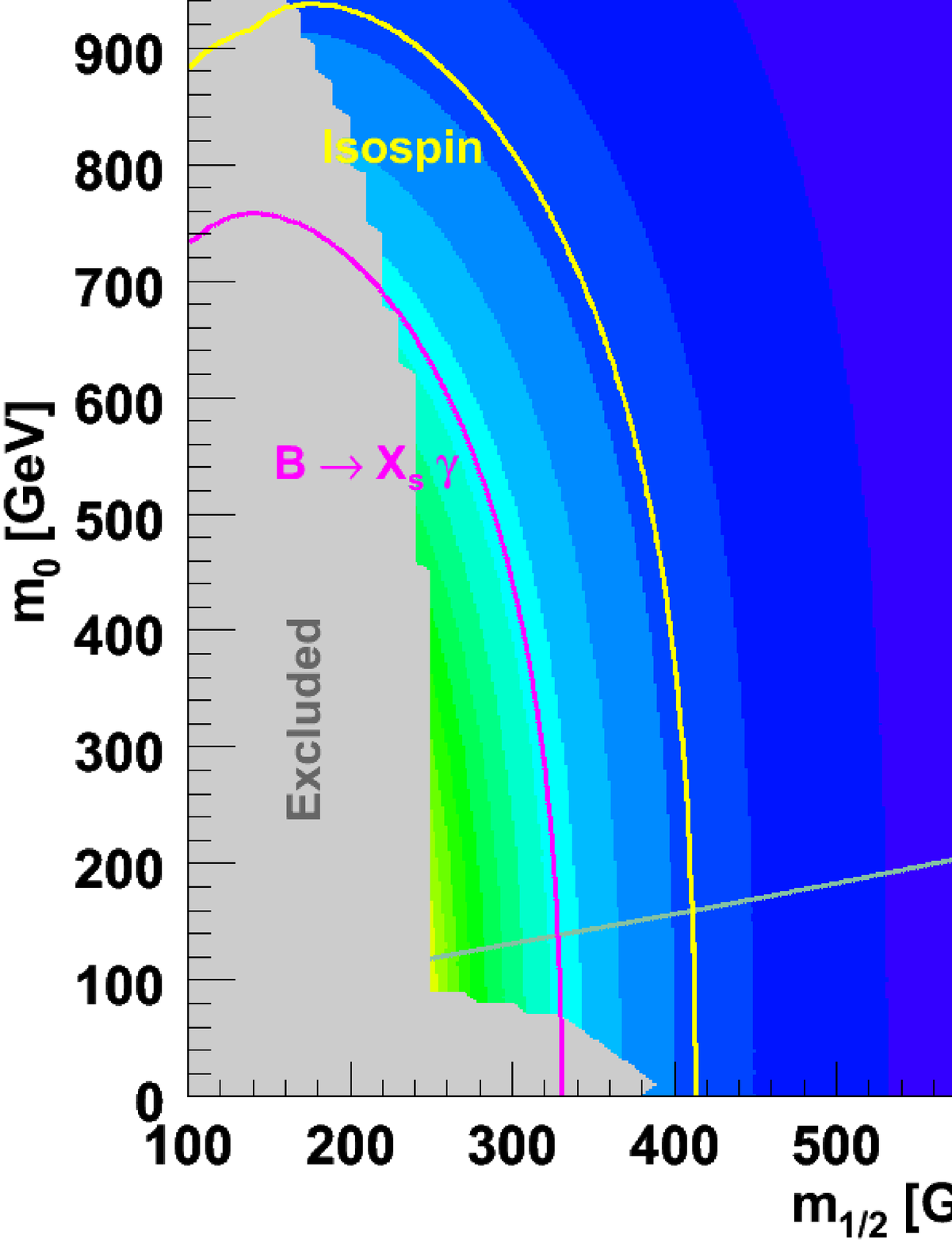}&\includegraphics[width=8cm,height=6.2cm]{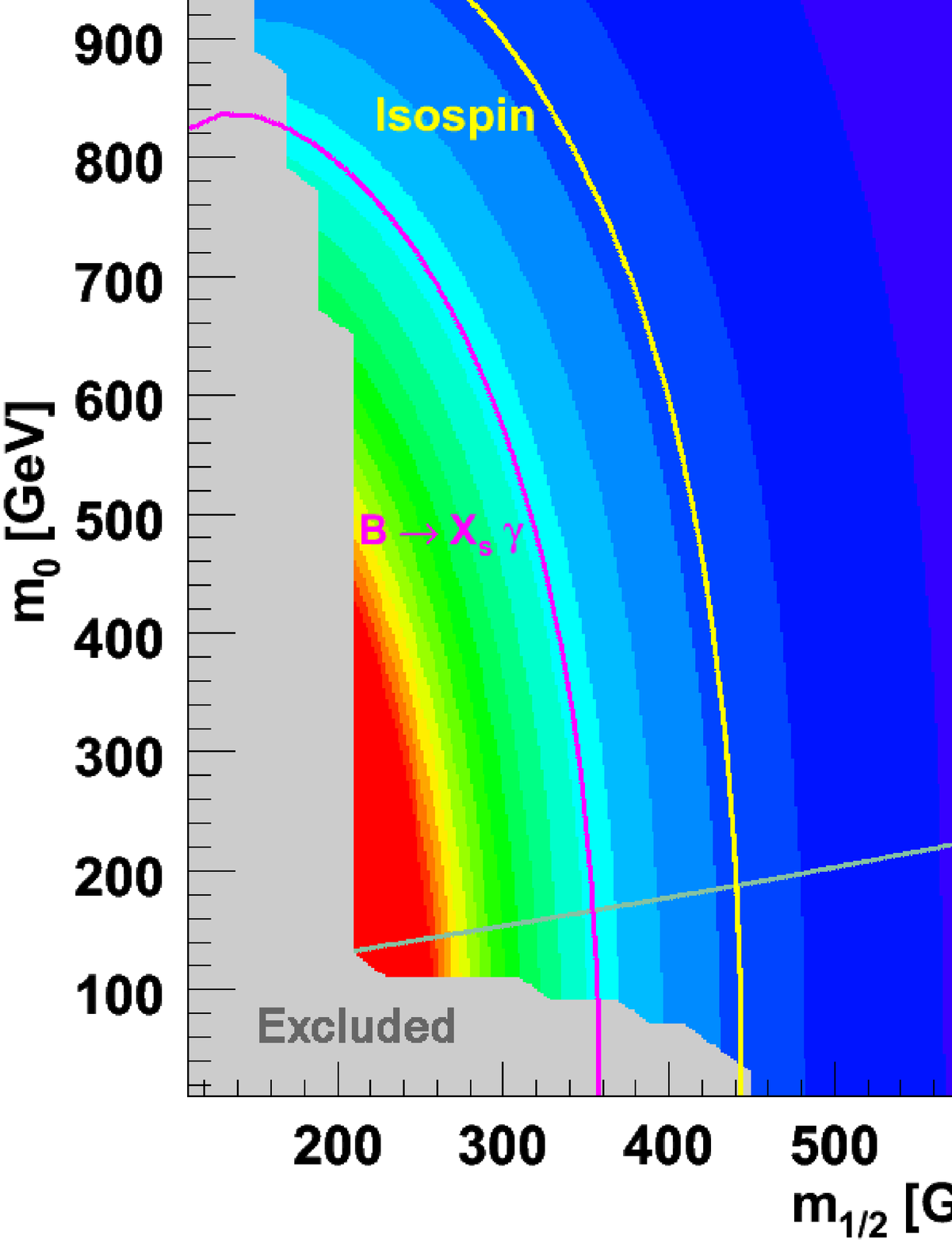}\\
\includegraphics[width=8cm,height=6.2cm]{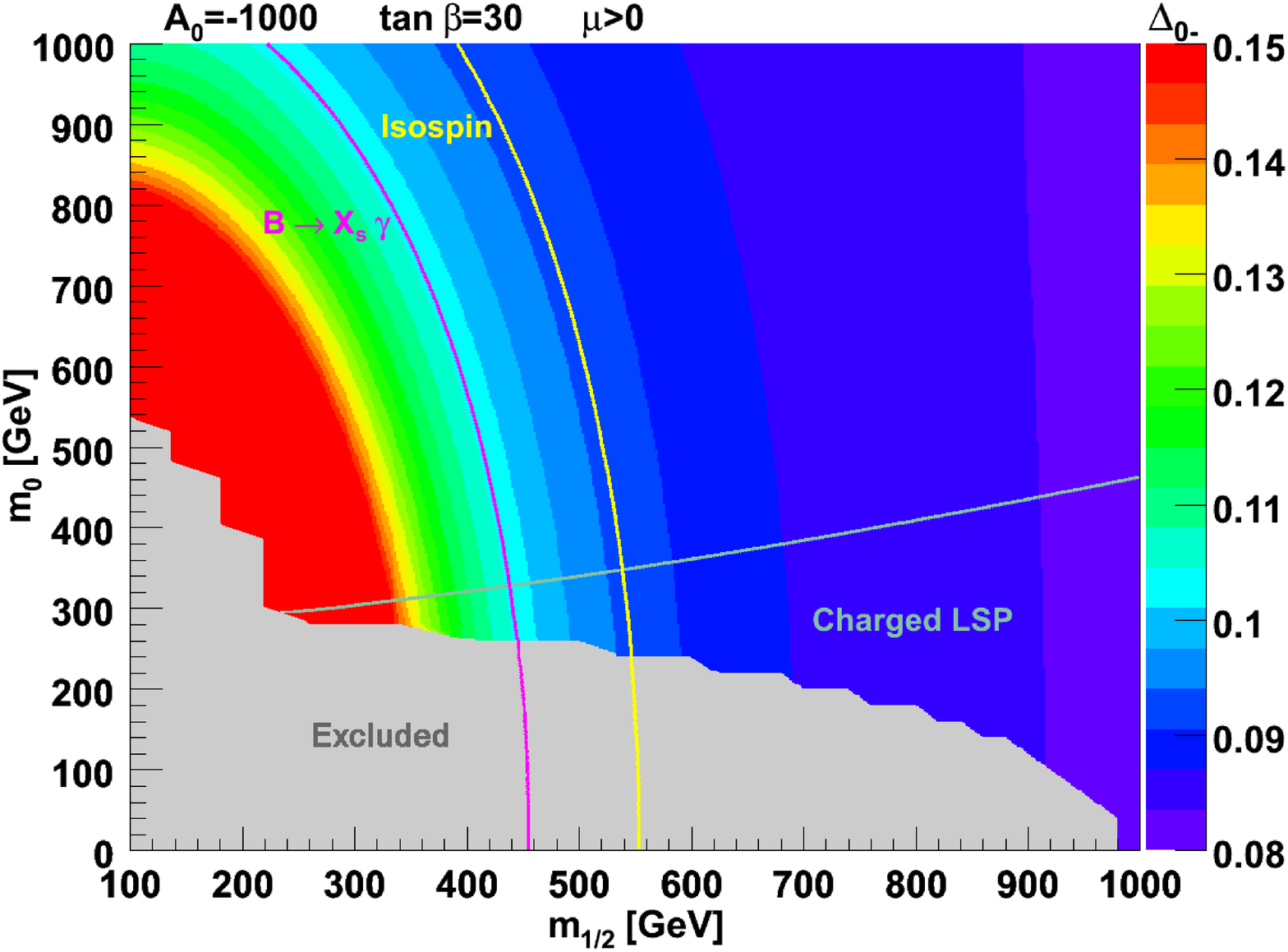}&\includegraphics[width=8cm,height=6.2cm]{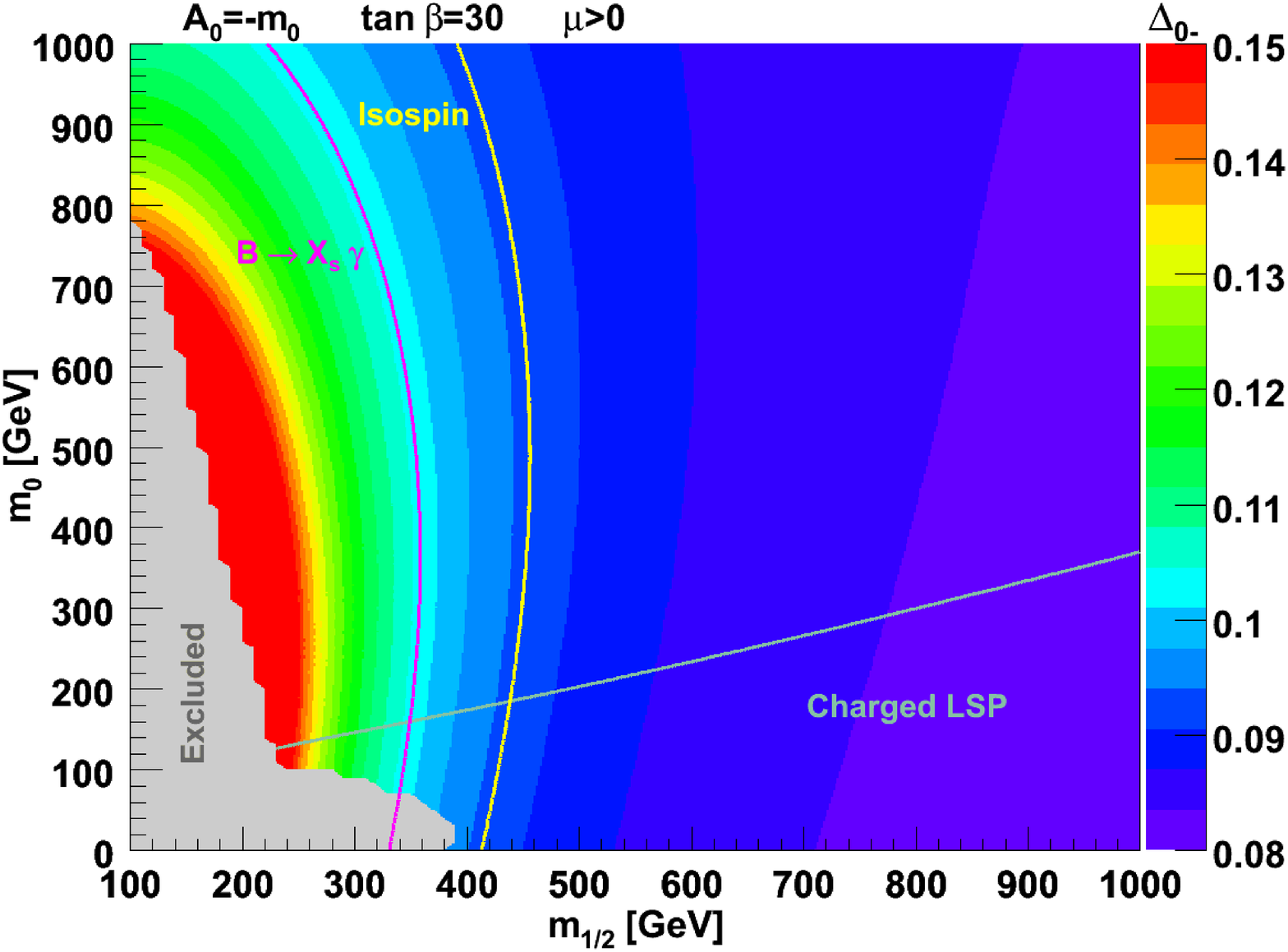}
\end{tabular}
\caption{Constraints on the mSUGRA parameter plane $(m_{1/2},m_0)$ for $\tan\beta=30$ and for several values of $A_0$. The definitions of the different regions are given in the text.}
\label{A0}
\end{figure}\\
An investigation of the $(m_{1/2},m_0)$ plane for $A_0=500,\,200,\,0,\,-200,\,-1000$ and $A_0=-m_0$, and for $\tan\beta=30$, is presented in Figure~\ref{A0}. In this figure, the area marked ``Isospin'' corresponds to the region excluded by the isospin breaking constraints, whereas the area marked ``$B\to X_s\gamma$'' corresponds to the region excluded by the inclusive branching ratio constraints. The ``Excluded'' area corresponds to the case where at least one of the particle masses does not satisfy the constraints of Ref. \cite{PDG2004}. And finally, ``Charged LSP'' is the cosmologically disfavored region. The various colors represent the intensity of the isospin asymmetry.\\
We can note here that the isospin symmetry breaking provides stringent constraints on the parameter space. It is more sensitive to the smaller values of $A_0$, and one can remark that it also decreases with larger $m_0$ and $m_{1/2}$.  Furthermore, it appears that the constraints from isospin asymmetry are more stringent than the ones from inclusive branching ratio. The results of Fig.~\ref{A0} concern a fixed value for $\tan\beta$. However, the isospin asymmetry increases with $\tan\beta$ and for example, for $A_0=500$, we can obtain valuable constraints at $\tan\beta=40$. This behaviour is revealed in Fig.~\ref{tanb}, where the isospin asymmetry is depicted in function of $\tan\beta$ and $m_{1/2}$ for $A_0=0$ and $A_0=-m_0$. We note that again, isospin asymmetry reveals to be more restricting than the branching ratio.\\
To conclude, the isospin asymmetry appears as a promissing observable to constrain the supersymmetric parameter space.

\begin{theacknowledgments}
F.M. acknowledges the support of the McCain Fellowship at Mount Allison University and a partial support from NSERC, and would also like to thank the organizers of SUSY06. M.A.'s research is partially funded by a discovery grant from NSERC.
\end{theacknowledgments}

\end{document}